\documentclass[a4paper]{article}
\usepackage[latin1]{inputenc}
\usepackage[english]{babel}
\usepackage{graphicx}
\usepackage[namelimits]{amsmath}
\usepackage{amssymb} 
\usepackage{amsmath} 
\usepackage{amsthm}

\newcommand{\vb}[1]{{\mbox{\boldmath$#1$}}} 

\newcommand{\x}{\xi}

\newcommand{\e}{\epsilon}
\newcommand{\xb}{\bar{\xi}}
\newcommand{\nb}{\bar{n}}
\newcommand{\Nb}{\bar{N}}
\newcommand{\eb}{\bar{\epsilon}}
\newcommand{\Gb}{\bar{G}}

\begin{document}

\setlength{\baselineskip}{0.7cm}

\title{On the probabilistic approach for Gaussian Berezin integrals}

\author{Massimo Ostilli
\\
Dipartimento di Fisica, Universit\`a di Roma ``La Sapienza'',\\ 
Piazzale A. Moro 2, Roma 00185, Italy}
\date{\today} \maketitle

\begin{abstract}
We present a novel approach to Gaussian Berezin correlation functions.
A formula well known in the literature expresses these quantities
in terms of submatrices of the inverse matrix appearing in the 
Gaussian action.
By using a recently proposed method to calculate Berezin integrals 
as an expectation of suitable functionals of Poisson processes,
we obtain an alternative formula which allows one to skip the 
calculation of the inverse of the matrix.
This formula, previously derived using different approaches  
(in particular by means of the Jacobi identity for the compound 
matrices), has computational advantages which grow rapidly with 
the dimension of the Grassmann algebra and the order of correlation. 
By using this alternative formula, we establish a mapping between 
two fermionic systems, not necessarily Gaussian, with short and long 
range interaction, respectively.
\end{abstract}


\section{Introduction}

The concept and the use of anticommuting variables originates both in the contest of the functional integral
approach to the quantization of fermionic systems \cite{B1,ZJ} and in several combinatorial problems (e.g. to represent the partition function of the planar Ising model) \cite{B2,B3}. The anticommuting character implies that these
variables belong to a Grassmann algebra.

A Grassmann algebra ${\cal G}_{g}$ of $g$ generators $\{\xi_{1}, \xi_{2}, \dots ,\xi_{g} \}$, 
is defined by the identity $\vb{1}^{g}$ and requiring the generators obey the following anticommutation relations: 
\begin{eqnarray}
\label{GRASMANN}
 \{ \xi_{i},\xi_{j} \} = \xi_{i} \xi_{j} + \xi_{j} \xi_{i} = 0 \ ~~ \forall \  i,j \in \{1,\dots g\}.  
\end{eqnarray}
Elements of ${\cal G}_{g}$ of the form $\xi_{i_{1}}\xi_{i_{2}} \dots \xi_{i_{k}}$ with 
$1 \leq i_{1} \leq i_{2} \leq \dots \leq i_{k} \leq g$ and $k \leq g$, are called monomials. 
Each elements of ${\cal G}_{g}$, $F(\vb{\xi})$,
can be written as a unique complex linear combination of
the $2^{g}$ (independent) monomials:
\begin{eqnarray}
F(\vb{\xi})=f_{0}\vb{1}^{g}+
\sum_{k=1}^{g}~\sum_{1 \leq i_{1} \leq i_{2} \leq \dots \leq i_{k} \leq g}
f_{i_{1},i_{2} \dots ,i_{k}}\xi_{i_{1}}\xi_{i_{2}} \dots \xi_{i_{k}} \nonumber ,
\end{eqnarray}
where $f_{i_{1},i_{2} \dots ,i_{k}} \in \mathbb{C}$. 
In this algebra, derivation and integration (the Berezin integral) are properly defined.
The reader is referred to \cite{B1,D,FS} for details. 
Let us introduce the symbols $d\xi_{1}, \ldots, d\xi_{g}$ satisfying the following formal
relations
\begin{eqnarray} 
\label{FORMALRELATIONS}
 \{ d\x_{i},d\x_{i} \} = \{ d\x_{i},\x_{j} \} =0. 
\end{eqnarray}
Let us define the elementary Berezin integrals 
\begin{eqnarray} 
 \int^{B} d\x_{i} = 0 , \nonumber
\mbox{ and for } j \neq i ~
 \int^{B} \x_{j} \x_{i} d\x_{i} = \x_{j} ,
\end{eqnarray} 
iterating this relations we arrive to 
\begin{eqnarray}  
 \int^{B} \x_{1} \x_{2}\ldots \x_{g} d\x_{1}...d\x_{g} = 1,
 \nonumber
\end{eqnarray}  
so, from linearity we have the following general definition of Berezin integral 
\begin{eqnarray}
\label{BEREZIN}
\int^{B}F(\vb{\xi})d\xi_{1} d\xi_{2} \dots d\xi_{g} =f_{1, \dots , g}.
\end{eqnarray}
The derivation is defined as 
\begin{eqnarray} 
\label{DERIVATION}
 {\delta \over {\delta  \xi_{i}}} \xi_{\mu_{1}}\cdots \xi_{\mu_{n}} & \equiv & 
 \delta_{\mu_{1}i} \xi_{\mu_{2}}\xi_{\mu_{3}}\cdots \xi_{\mu_{n}}-\delta_{\mu_{2}i}
 \xi_{\mu_{1}}\xi_{\mu_{3}}...\xi_{\mu_{n}}+... \nonumber \\ 
 && +(-1)^{n-1}\delta_{\mu_{1}n} \xi_{\mu_{1}} \xi_{\mu_{2}}\cdots \xi_{\mu_{n-1}}.
\end{eqnarray}

Let now ${\cal G}_{2g}$ be a Grassmann algebra over $\mathbb{C}$, with $2g$ generators, which we label as $g$ barred and $g$ unbarred: 
$\xi_{1}, \xi_{2}, \dots ,\xi_{g} ; \bar{\xi}_{1}, \bar{\xi}_{2}, \dots ,\bar{\xi}_{g}$. 
Let $\mathbb{Z}_{2}^{*}= \{ 0,1 \} $ and $\mathbb({Z}_{2}^{*})^{\times g} =\mathbb{Z}_{2}^{*} \otimes
\mathbb{Z}_{2}^{*} \otimes ...  \otimes \mathbb{Z}_{2}^{*} $ (the direct product of $\mathbb{Z}_{2}^{*}$ $g$-times).
Let us define the normalized Gaussian Berezin integrals: 
\begin{eqnarray}
\label{GAUSS}
\int^{B} \vb{\psi^{s} d\mu} \equiv
{{ \int^{B} \exp (- \vb{\xb, A \x}) \x_{1}^{n_{1}}\xb_{1}^{\nb_{1}} \ldots \x_{g}^{n_{g}}\xb_{g}^{\nb_{g}}   d\x_{1}d\xb_{1}...d\x_{g}d\xb_{g}} \over { \int^{B} \exp (-\vb{\xb,A\x})   d\x_{1}d\xb_{1}...d\x_{g}d\xb_{g}} }
\end{eqnarray} 
where $\vb{A}$ is any invertible $g \times g$ matrix, 
$\vb{\psi}=(\x_{1},\xb_{1},...,\x_{g},\xb_{g})$,
$\vb{\psi^{s}}= \x_{1}^{n_{1}}\xb_{1}^{\nb_{1}} 
\ldots \x_{g}^{n_{g}}\xb_{g}^{\nb_{g}}$ with
\begin{equation}
\vb{s}=(n_{1},\nb_{1},...,n_{g},\nb_{g}) \in ({\mathbb{Z}_{2}}^{*})^{2g}
\label{esse}
\end{equation} 
and the definition of the Gaussian measure $\vb{d\mu}$ is implicit in (\ref{GAUSS}).
Let $\bar{k}$ be the number of labels $i$ such that $\nb_{i}=1$ and $k$ the number of
 labels $j$ such that $n_{j}=1$ .
Due to the parity of the Gaussian function $\exp (-\vb{\xb,A\x})$, it is easy to see that the integral (\ref{GAUSS}) vanishes if $k \neq \bar{k}$.

When $k=\bar{k}$, the integral (\ref{GAUSS}) can be evaluated by a well known formula in terms of the determinant of a submatrix of the inverse matrix $\vb{C=A}^{-1}$ (see e.g. \cite{ZJ}):
\begin{eqnarray}
\label{SOLUZ1}
\int^{B} \vb{\psi^{s} d\mu} = \det \vb{C_{s}},
\end{eqnarray}
where $\vb{C_{s}}$ is the matrix obtained from $\vb{C}$ erasing the rows with label $j$ if $n_{j}=0$ and the columns with label $i$ if $\nb_{i}=0$. If $\vb{s}=(0, \dots ,0)$, we define $\vb{C_{s}}=1$.
Equation (\ref{SOLUZ1}) is then expressed in terms of a $k \times k$ submatrix of the inverse of $\vb{A}$.

Starting from a recently developed approach to calculate Berezin integrals \cite{DJS}, we will derive (see Eq. (\ref{SOLUZ2}))
a formula for (\ref{GAUSS}) 
in terms of the determinant of a $(g-k) \times (g-k)$ sub matrix of $\vb{A}$. When compared with the (\ref{SOLUZ1}),
this formula gives rise to an algebraic identity, which is nothing else than Jacobi's theorem relating the minors 
of a matrix to the minors of the inverse of the same matrix, see e. g. \cite{AITKEN,GANTMACHER}, see also \cite{KLAUS}. 
Although this identity and the Gaussian Berezin correlation functions can be derived algebraically in a simpler manner, even not using the Jacobi's theorem \cite{KERLER,CHARRET,SORIN}, a probabilistic derivation provides an important example of how the stochastic approach for fermions models can be applied to obtain exact results. 

It is worth considering that even if this identity is well
known among mathematicians, it seems that it has been rarely explored 
in Quantum Field Theory \cite{KERLER,CHARRET,KLAUS}. 
In section 5 we will explain some interesting aspects and physical applications of this identity.
In particular we will find a mapping between two fermionic systems. According to this mapping  
the Grassmannian partition function of a short range system having $\vb{A}$ as matrix 
for the free (Gaussian) part of the action, equals the Grassmannian partition function of a long range system
having $\vb{A}^{-1}$ as matrix for the free part.

In the remainder we will give the proof providing in full detail all the calculations
as they may look unfamiliar to many readers.
\section{Gaussian Berezin integrals}

Given $\vb{s}=(n_{1},\nb_{1},...,n_{g},\nb_{g}) \in ({\mathbb{Z}_{2}}^{*})^{2g}$ , let 
\begin{eqnarray}
\vb{s^{*}}=(1-\nb_{1},1-n_{1},...,1-\nb_{g},1-n_{g}).
\end{eqnarray}  
Probabilistically, we will show that besides 
\begin{eqnarray}
\label{I0}
{ \int^{B} \exp (- \vb{\xb, A \x}) \vb{\psi}^{\vb{s}}  d\x_{1}d\xb_{1}...d\x_{g}d\xb_{g}}=\det \vb{C_{s}}\det \vb{A} 
\end{eqnarray}
holds also
\begin{equation}
\label{I}
{ \int^{B} \exp (- \vb{\xb, A \x}) \vb{\psi}^{\vb{s}}  d\x_{1}d\xb_{1}...d\x_{g}d\xb_{g}}=\det \vb{A_{s^{*}}}(-1)^{W(\vb{s})}. 
\footnote{According to the definition given for $\vb{C_{s}}$ in introduction, $\vb{A_{s^{*}}}$ is the $(g-k) \times (g-k)$ submatrix of $\vb{A}$ obtained from $\vb{A}$ 
erasing the rows with label
$i$ if $\nb_{i}=1$ and the columns with label $j$ if $n_{j}=1$. 
If $\vb{s}=(1, \dots ,1)$ we define $\vb{A_{s^{*}}}=1$.}
\end{equation}
The factor$(-1)^W(\vb{s})$ determines a global sign with
\begin{eqnarray}
\label{SIGN}
W(\vb{s}) = (n_{1} + \nb_{1}) + 2(n_{2} + \nb_{2}) + \dots + g(n_{g} + \nb_{g}) 
\end{eqnarray}     
From the above equations we get the following expression for 
the normalized Gaussian Berezin integrals 
\begin{eqnarray}
\label{SOLUZ2}
\int^{B} \vb{\psi^{s} d\mu}={\det \vb{A_{s^{*}}} \over {\det \vb{A}}} (-1)^{W(\vb{s})}, 
\end{eqnarray}
and by using Eq. (\ref{SOLUZ1}) the Jacobi's identity follows 
\footnote{It is simple to observe that $(n_{1} + \nb_{1}) + 2(n_{2} + \nb_{2}) + \dots + g(n_{g} + \nb_{g})=
i_{1}+j_{1}+i_{2}+j_{2}+ \dots i_{k}+j_{k}$, where $i_{1}\dots i_{k}$ and $j_{1} \dots j_{k}$ are the the sets of indices
such that $n_{i_{\nu}}=1$ and $n_{j_{\nu}}=1$ respectively, for $\nu=1 \dots k$.
In this form the sign in the Jacobi's relation is normally presented in literature.}
\begin{eqnarray}
\label{IDENTITY}
{\det \vb{A_{s^{*}}} \over {\det \vb{A}}} (-1)^{W(\vb{s})}=\det \vb{C_{s}}.
\end{eqnarray}
\section{Berezin integrals as averages over Poisson processes}

Let $\mathbb{Z}_{2}= \{ -1,1 \} $ and $\mathbb{Z}_{2}^{\times g} =\mathbb{Z}_{2} \otimes
\mathbb{Z}_{2} \otimes ...  \otimes \mathbb{Z}_{2} $ (the direct product of $\mathbb{Z}_{2}$ $g$-times). 
Let $\vb{\e}$ and $\vb{\sigma} \in \mathbb{Z}_{2}^{\times g}$ and $\vb{e}_{g}=(1,\dots,1) \in \mathbb{Z}_{2}^{\times g}$. 
Let $\{ N_{t}^{\vb{\e}} \}_{\vb{\e} \neq \vb{e}_{g}}$ be a family of $2^{g}-1$ left-continuous 
independent Poisson processes of pure growth with
unit parameter. These Processes are then characterized by the probabilities
\begin{eqnarray}
\label{POISSON}
P(  N_{t+\Delta t}^{\vb{\e}} - N_{t}^{\vb{\e}}=k)=\frac{(\Delta t)^{k}}{k!}\exp{(-\Delta t)}
\end{eqnarray} 

Following \cite{DJS} \footnote{ In \cite{JP}
it is shown in a simple manner, how to derive the Poissonian 
probabilistic representation of the time evolution operator
for any operator described by a finite matrix. This derivation can be used instead of that
appearing in the fifth chapter of \cite{DJS}. }, 
given any ``action'' $S(\vb{\xi}) \in {\cal G}_{g}$, we can calculate the Berezin integrals by an expectation
of a suitable functionals of Poisson processes using the following formula:
\begin{eqnarray}
\label{CORRELATION2}
\int^{B}\x_{1}^{n_{1}}  &\ldots& \x_{g}^{n_{g}} \exp(-S(\vb{\x}))d\x_{1}\dots d\x_{g} = 
\Delta_{g}(-(-1)^{{\vb{n}}}) \exp \bigl( |\Gamma|-s({\vb{e}_{g}}) \bigr) \nonumber \\ &&
\times \vb{E} \biggl( \frac{ \prod_{l=1}^{g}1-(-1)^{{N_{1}^{l}+n_{l}}} } {2} \nonumber \\ && 
 \times \exp \biggl\{ \sum_{\vb{\e} \in \Gamma} \int_{[0,1)} \log (-s(\vb{\e})
C_{g}(\vb{\e},-(-1)^{\vb{N_{s}+n}}))dN_{s}^{\vb{\e}} \biggr\} \biggr), 
\end{eqnarray}
where: 
\begin{eqnarray}
\label{DELTA}
\Delta_{g}(\vb{\sigma})= \prod_{l=1}^{g} \biggl( {{1-\sigma_{l}} \over 2}+ {{1+\sigma_{l}} \over 2}
 \sigma_{1} \ldots \sigma_{l-1} \biggr),  
\end{eqnarray}
\begin{eqnarray}
\label{C}
&& C_{g}(\vb{\epsilon,\sigma})=\prod_{l=1}^{g} \biggl( {{1+\e_{l}} \over 2} + {{1-\e_{l}} \over 2}
  {{1-\sigma_{l}} \over 2} \e_{1}\ldots \e_{l-1} \sigma_{1} \ldots \sigma_{l-1} \biggr), 
\end{eqnarray}
\begin{eqnarray} 
(-1)^{\vb{n}}=((-1)^{n_{1}},...,(-1)^{n_{g}})  \in  \mathbb{Z}_{2}^{\times g},
\end{eqnarray}
\begin{eqnarray}
\label{SPICCOLO}
 s(\vb{\e})&=&\Delta(\vb{\e})\int^{B}\vb{\x}^{{1 \over 2}(\vb{1+\e})}S(\vb{\x})d\x_{1}\dots d\x_{g} 
 , \mbox{where} \nonumber 
\\ \vb{\x}^{{1 \over 2}(\vb{1+\e})}&=&(\x_{1}^{{1 \over 2}^{(1+\e_{1})}} \dots \x_{g}^{{1 \over 2}^{(1+\e_{g})}}), 
\end{eqnarray}
\begin{eqnarray}
\label{GAMMA}
\Gamma = \{ \vb{\e} \in \mathbb{Z}_{2}^{\times g}:  \vb{\e} \neq \vb{e}_{g}, \ 
s(\vb{\e})C_{g}(\vb{\e,\sigma}) \neq 0 \}; \ |\Gamma |= \mbox{the cardinality of} \ \Gamma , 
\end{eqnarray}
\begin{eqnarray} 
\label{Nt}
\vb{N_{t}}=(N_{t}^{1},...,N_{t}^{g}); \ N_{t}^{l}=\sum_{\vb{\e} \in \Gamma}{1 \over 2}(1-\e_{l})N_{t}^\vb{\e}.
\end{eqnarray}
Finally the stochastic integrals that appear in (\ref{CORRELATION2}) are ordinary Stieltjes integrals:
\begin{eqnarray}
\label{STIELTJES}
\int_{[0,t)}f(s,N_{s})dN_{s}=\sum_{s_{h}<t}f(s_{h},N_{s_{h}}),
\end{eqnarray}
$s_{h}$ being the jump times of the process. 
We refer the reader to \cite{DJS,JP} for proofs and details. 
Here we want to emphasize that: 
\textit{only the trajectories such that  
$N_{1}^{\vb{\e}} \in \{0,1\},\forall \vb{\e} \in \Gamma$ 
can contribute to the expectation in (\ref{CORRELATION2})}.
We will indicate such a set as the set $''zero-one''$ .
This happens due to the structure of the coefficients $C_{g}$ present in the argument of 
the logarithm which occurs in Eq. (\ref{CORRELATION2}). In fact,
let us consider the expectation of the exponential occuring in (\ref{CORRELATION2}),
if we look at the definitions of the coefficients $C_{g}$, neglecting their moduli,
we can write this expectation
as a sum of terms corresponding to the following families of events with respective
contributions and probabilities (the latter are obtained by integrating over the jumps
times the infinitesimal probabilities (\ref{POISSON})):
\begin{itemize}
\item
event in which no process jumps. This event gives 
$1$ with probability $\exp(-|\Gamma|)$;
\item
events in which only one process, say with a generic label $\vb{\e}^{1}$, makes a jump. 
Each of these events will give the value
$-s(\vb{\e}^{1})\prod_{l:  \e_{l}^{1}=-1}{{1+(-1)^{n_{l}}} \over 2}$ with probability
$\int_{0}^{1}ds_{1}\exp(-|\Gamma|)=\exp(-|\Gamma|)$;
\item
events in which two and only two processes with different labels, say $\vb{\e}^{1}$
 and $\vb{\e}^{2}$, jump at 
times $s_{1}<s_{2}$ respectively. Each of these events will give the value   
$(-s(\vb{\e}^{1}))\prod_{l:  \e_{l}^{1}=-1}{{1+(-1)^{n_{l}}} \over 2 }\times  
(-s(\vb{\e}^{2}))\prod_{l:  \e_{l}^{2}=-1}{{1+(-1)^{n_{l}+ {1 \over 2}(1-\e_{l}^{1}) }} \over 2}$ with probability ${\int_{0}^{1}}_{s_{1} < s_{2}}ds_{1}ds_{2}\exp(-|\Gamma|)$
\item
and so on \ldots
\item
events in which  one single process, say with label $\vb{\e}^{1}$, jumps two times. Each of these events gives: 
$-s(\vb{\e}^{1})\prod_{l:  \e_{l}^{1}=-1}{{1+(-1)^{n_{l}}} \over 2}\times
(-s(\vb{\e}^{1}))\prod_{l:  \e_{l}^{1}=-1}{{1+(-1)^{n_{l}+1}} \over 2}$ with probability
${\int_{0}^{1}}_{s_{1} < s_{2}}ds_{1}ds_{2}\exp(-|\Gamma|)$
\item
and similarly for processes with more than two jumps.
\end{itemize}
It is then clear that the last type of contributions gives always zero.
Furthermore note that events in which two or more processes jump at the same time, have probability zero,
even if $C_{g} \neq 0$. 

So, given  $S(\vb{\x})$ and then $s(\vb{\e})$, one will have to determine what is the set of the effective trajectories, in general
a smaller subset of the above set of trajectories $zero-one$. In the next section we will investigate the Gaussian case 
($S$ bilinear in the $\x$) and we will see that such a set is ``very small''. 
The search for the effective trajectories and their phases will allow us to automatically construct the matrix $\vb{A_{s^{*}}}$
and, in the final expression, to recover its determinant up to a global sign.
\section{Proof of equation (\ref{SOLUZ2})}

The application of Eq. (\ref{CORRELATION2}) will give us the Gaussian integral not normalized,
let be $I$ (the numerator of (\ref{GAUSS})).
Let us consider (\ref{CORRELATION2}) with $g$ barred variables and $g$ unbarred variables.
We will have to calculate various parts.
In section 4.1 we will calculate $s(\vb{\e})$ and then we will have $\Gamma$ for our bilinear case,
 next we will calculate partially $C_{2g}$. This result will be used in section 4.2 for calculating the effective 
trajectories and in section 4.3 for evaluating the phase of (\ref{CORRELATION2}) which
will provide us the final formula. 
We will consider the integral
with respect to $d\xb_{1}...d\xb_{g}d\x_{1}...d\x_{g}$ (instead of $d\x_{1}d\xb_{1}...d\x_{g}d\xb_{g}$).
Therefore, we will evaluate the following integral
\begin{eqnarray}
\label{NUMERATORTILDE}
 \widetilde{I}=\int^{B}d\xb_{1}...d\xb_{g}d\x_{1}...d\x_{g}\xb_{1}^{\nb_{1}}\ldots
 \xb_{g}^{\nb_{g}}\x_{1}^{n_{1}}\ldots \x_{g}^{n_{g}} \exp(-\vb{\xb,A\x}),
\end{eqnarray}
which is related to $I$ by a simple correcting sign, $I=\widetilde{I}(-1)^{Q}$
which we will calculate in section 4.3. We will see that $Q$ depends on $g$ and $\vb{s}$.
With the chosen ordering of the generators of the Grassmann algebra, $\vb{\e}$ and $\vb{s}$
should be changed into
\begin{eqnarray}
\label{epsNEW}
\widetilde{\vb{\e}}=(\eb_{1},...\eb_{g};\e_{1},...,\e_{g}), 
\end{eqnarray}
\begin{eqnarray}
\label{sNEW}
 \widetilde{\vb{s}} =(\nb_{1},...,\nb_{g};n_{1},...,n_{g}).
\end{eqnarray}
However, for simplicity, we will continue to use, if not ambiguous, the initial notations
$ \ \vb{\e}$ and $\vb{s}$ also for the new arrangement of the Grassmann generators.
At the end we will return to the true integral corresponding to the original arrangement
of the Grassmann generators by using $I=\widetilde{I}(-1)^{Q}$. 

\subsection{Calculation of the factors: $s(\vb{\e}), \ \Delta_{2g}(\vb{\e})$ and $C_{2g}$}
Let us start by calculating $s(\vb{\e})$ defined by Eq. (\ref{SPICCOLO}):
\begin{eqnarray}
\label{SPICCOLOG} 
&& s(\vb{\e})=\Delta_{2g}(\vb{\e})\int^{B}d\xb_{1}...d\xb_{g}d\x_{1}...d\x_{g}\xb_{1}^{{1 \over 2}(1+\eb_{1})}
\ldots \xb_{g}^{{1 \over 2}(1+\eb_{g})}\x_{1}^{{1 \over 2}(1+\e_{1})}\ldots \x_{g}^
{{1 \over 2}(1+\e_{g})}(\vb{\xb,A\x}) 
\nonumber \\ &&= \Delta_{2g}(\vb{\e})\int^{B}d\xb_{1}...d\xb_{g}d\x_{1}...d\x_{g} \xb_{1}^{{1 \over 2}(1+\eb_{1})}
\ldots \xb_{g}^{{1 \over 2}(1+\eb_{g})}\x_{1}^{{1 \over 2}(1+\e_{1})}\ldots \x_{g}^
{{1 \over 2}(1+\e_{g})}\sum_{ij}\xb_{i},A_{ij}\x_{j}  
\nonumber \\ && =
\Delta_{2g}(\vb{\e})\sum_{ij}A_{ij}(-1)^{p_{ij}}\int^{B}d\xb_{1}...d\xb_{g}d\x_{1}...d\x_{g}  \xb_{1}^{{1 \over 2}(1+\eb_{1})}
\ldots \xb_{i}^{{1 \over 2}(1+\eb_{i})+1}\ldots \xb_{g}^{{1 \over 2}(1+\eb_{g})}
\nonumber \\ && \ \ \ \ \ \ \ \ \ \ \ \ \ \ \ \ \ \ \ \ \ \ \ \ \ \ \ \ \ \ \ \ \  \times
\ \x_{1}^{{1 \over 2}(1+\e_{1})}\ldots \x_{j}^{{1 \over 2}(1+\e_{j})+1}\ldots \x_{g}^
{{1 \over 2}(1+\e_{g})},
\end{eqnarray}
where $(-1)^{p_{ij}}$ accounts for the exchanges needed to bring $\xb_{i}$
in front of $\xb_{i}^{{1 \over 2}(1+\eb_{i})}$ and $\x_{j}$ in front of $\x_{j}^{{1 \over 2}(1+\e_{j})}$.
We see, that integral (\ref{SPICCOLOG}) vanishes unless $\vb{\e}$ has the form
\begin{eqnarray} 
\label{EPSILONG} 
 \vb{\e}_{ij}=(\eb_{1},...\eb_{g};\e_{1}...,\e_{g})
 \mbox{, with } \eb_{i}=\e_{j}=-1 \mbox{ and } \eb_{l}=\e_{l}=1 \mbox{ for } l \neq i,j.
\end{eqnarray}
Furthermore it follows from this that
\begin{eqnarray} 
 s(\vb{e}_{2g})=0, \mbox{where } \vb{e}_{2g}=(1,...,1;1....,1). \nonumber
\end{eqnarray}
The set $\Gamma$ is then characterized. Its elements are in the set of the $2g-ple$
$\e_{ij}$ such that $A_{ij} \neq 0$. In this way we can arrange the set 
of Poisson processes in a $g \times g$ matrix:
\begin{eqnarray}
\vb{\cal N}=((N_{s}^{ij})) \mbox{, where } N_{s}^{ij} \equiv N_{s}^{\e_{ij}}.   
\end{eqnarray}
Even if not explicitly indicated, all the sums and products will have to
be considered in the range such that $A_{ij} \neq 0$.

From Eq. (\ref{SPICCOLOG}) we have
\begin{eqnarray}
s(\vb{\e}_{ij})=\Delta_{2g}(\e_{ij})(-1)^{p_{ij}}A_{ij}.
\end{eqnarray}
In appendix B it is easily proved that one has
\begin{eqnarray}
s(\vb{\e}_{ij})=A_{ij}.
\end{eqnarray}
As regards the calculation of $C_{2g}$, 
according to our choice of the ordering of the Grassmann generators, we must consider (see Eq. (\ref{Nt}))
$\widetilde{\vb{N}}_{s}=(\vb{\Nb}_{s} ;\vb{N}_{s})$ and $\widetilde{\vb{n}}=(\vb{\nb ;n})$, 
where
\begin{eqnarray}
\label{NROWS}
\bar{N}_{s}^{l}=\sum_{\vb{\e} \in \Gamma}{1 \over 2}(1-\eb^{l})N^{\vb{\e}}_{s};
\end{eqnarray}
\begin{eqnarray}
\label{NCOLUMNS}
N_{s}^{l}=\sum_{\vb{\e} \in \Gamma}{1 \over 2}(1-\e^{l})N^{\vb{\e}}_{s}.
\end{eqnarray}
So to calculate $C_{2g}$ one has to evaluate
\begin{eqnarray}
 C_{2g}(\vb{\e}_{ij},-(-1)^{\widetilde{\vb{N}_{s}}+\widetilde{\vb{n}}})=C_{2g}\biggl( \vb{\e}_{ij},-\bigl(
 (-1)^{\Nb_{s}^{1}+\nb_{1}},...,(-1)^{\Nb_{s}^{g}+\nb_{g}};(-1)^{N_{s}^{1}+n_{1}},...,
 (-1)^{N_{s}^{g}+n_{g}} \bigr) \biggr).
\nonumber
\end{eqnarray}
In appendix C it is easily proved that one has
\begin{eqnarray}
&& C_{2g}(\vb{\e}_{ij},-(-1)^{\widetilde{\vb{N}_{s}}+\widetilde{\vb{n}}})={1+(-1)^{\Nb_{s}^{i}+\nb_{i}} \over 2}
 {1+(-1)^{N_{s}^{j}+n_{j}} \over 2} (-1)^{g+j+i+1} \nonumber \\ && \times
\exp \{i \pi [{\Nb_{s}^{i}+...+\Nb_{s}^{g}+N_{s}^{1}+...+N_{s}^{j-1}+\nb_{i}+...+\nb_{g}+n_{1}+...+n_{j-1}}] \} ,
\nonumber
\end{eqnarray}
where
\begin{eqnarray}
\label{N&N}
 \Nb_{s}^{i}=\sum_{l=1}^{g}N_{s}^{il} \nonumber \\
 N_{s}^{j}=\sum_{l=1}^{g}N_{s}^{lj}. 
\end{eqnarray}  
Therefore, as far as the exponential factor in (\ref{CORRELATION2}) is concerned, a given trajectory will give the
following contribution 
\begin{eqnarray}
\label{PRODUCT}
&& \prod_{ij} \exp \biggl[ \int_{[0,1)} \log(-A_{ij}) dN_{s}^{ij}\biggr] \exp \biggl[ \int_{[0,1)} \log [ {1+(-1)^{\Nb_{s}^{i}+\nb_{i}} \over 2}{1+(-1)^{N_{s}^{j}+\nb_{j}} \over 2} ]dN_{s}^{ij}\biggr] \nonumber \\ && \times 
\exp \biggl[ \int_{[0,1)} i \pi [g+i+j+1+ \sum_{m=i}^{g}(\Nb_{s}^{m}+\nb_{m})+
\sum_{m=1}^{j-1}(N_{s}^{m}+n_{m}) ] dN_{s}^{ij} \biggr], 
\end{eqnarray}  
where the range of the product is over the couples $ij$ with $A_{ij} \neq 0$.
\subsection{Effective trajectories}

In section 3, we have seen that the set of the effective trajectories is a subset of set $zero-one$
 (i.e. the set in which one single process can make at most
one jump during the time interval $[0,1)$) and we have anticipated that in the Gaussian case such a set is small.    
Let us concentrate on the second factor of the product (\ref{PRODUCT}).
Let us define:
\begin{eqnarray}
\label{GG}
 \Gb_{i} \cdot G_{j} = {1+(-1)^{\Nb_{s}^{i}+\nb_{i}} \over 2}~{1+(-1)^{N_{s}^{j}+n_{j}} \over 2}
\end{eqnarray}   
and remember that the processes are left continuous 
\footnote{This implies that if $s=s^{ij}$ is the jump time of the first jump for the process $N_{s}^{ij}$,
 then $N_{s}^{ij}=0$ for $s \leq s^{ij}$ and  $N_{s}^{ij}=1$ for $s > s^{ij}$.}. 
Let us note from (\ref{N&N}) that
$\Nb_{s}^{i}$
 and
$N_{s}^{j}$
 are the sums of the elements of the $i$th row and the
$j$th column of the matrix $\vb{\cal N}$ respectively.

Let us now consider the trajectories in which in every row and in every column of $\vb{\cal N}$ there is
at most just one process which jumps (clearly, one process which jumps in some row, jumps
in some corresponding column too). 
With such events, if for example $N_{s}^{ij}$ is a process which jumps at time $s^{ij}$, then according to Eq. (\ref{GG})
we have:
\begin{eqnarray}
 \Gb_{i} \cdot G_{j}|_{s=s^{ij}} ={1+(-1)^{\nb_{i}} \over 2}~{1+(-1)^{n_{j}} \over 2}.
\end{eqnarray}   
Since we consider the factor corresponding to the link $(ij)$ and carry out the stochastic integral 
in $dN_{s}^{ij}$, we see that in order it does not give $\exp \log (0)$, it must be $\nb_{i}=n_{j}=0$.

Let us now consider one event in which there is at least one row in which at least two processes,
$N_{s}^{ij}$ and $N_{s}^{ik}$, jump, at times $s^{ij}$ and $s^{ik}$ respectively. Let us consider the ($ij$) factor in Eq. (\ref{PRODUCT}). We have to integrate in $dN_{s}^{ij}$, i.e. we have to evaluate the integrand at time $s^{ij}$.
Let $\nb_{i}=1$. We must consider two possibilities:
\begin{enumerate}
\item if $s^{ij}<s^{ik}$ we have: \\
$\Nb_{s^{ij}}^{i}=0$ from which follows $\Gb_{i}=0$  
\item if $s^{ij}>s^{ik}$ we have: \\
 $\Nb_{s^{ij}}^{i}=1$ from which follows $\Gb_{i}=1$.
\end{enumerate}
Therefore the stochastic integral can be finite in the the case 2, but as we will integrate in 
$dN_{s}^{ik}$, we will have to calculate the integrand at time
$s^{ik}$ then obtaining $\Gb_{i}=0$.
Finally Let us note that the event $s^{ij}=s^{ik}$ has zero probability.

Let $\nb_{i}=0$. Repeating a similar argument we will arrive at the same conclusions.
Of course, the same discussion can be done for the columns. 

We have proved that the effective trajectories belong to the set in which in any row and in any column of the matrix 
$\vb{\cal N}$ there is at most one jumping process. 
Let us call such a set:
\begin{eqnarray}
{\cal C}_{1} &=& \{ \mbox{set of trajectories in which for any row and for any column}
\nonumber \\ &&  \mbox{of the matrix $\vb{\cal N}$ there is at most one jumping process} \} . 
\end{eqnarray}
Furthermore we have proved that as $\nb_{i}=1$ and/or $n_{j}=1$, then it must be $N_{1}^{ij}=0$. 
Let us call such set:  
\begin{eqnarray}
 {\cal C}_{2}&=& \{ \mbox{set of the trajectories in which if } \nb_{i} \mbox{ and/or } n_{j}=1 \nonumber \\
 && \mbox{ then } N_{1}^{ij}=0 \} . 
\end{eqnarray}
If now we consider (\ref{PRODUCT}), (regardless of the phase), imposing these restrictions by using
the product of two characteristic functions, we can eliminate the stochastic integral writing (\ref{PRODUCT}) as:
\begin{eqnarray}
\label{PRODUCT1}
 \prod_{ij}(-A_{ij})^{N_{1}^{ij}}\cdot \chi(  {\cal C}_{1} )\chi(  {\cal C}_{2} ){\cal S}_{ij},
\end{eqnarray}
where with ${\cal S}_{ij}$ we have indicated the phase (i. e.  the last of the three factors appearing 
in (\ref{PRODUCT})). 
Observing Eq. (\ref{CORRELATION2}), we see that to complete the search of the effective trajectories, we must multiply the above product by the following factor
\begin{eqnarray}
 \bar{F} \cdot F = \prod_{l=1}^{g}{{1-(-1)^{\Nb_{1}^{l}+\nb_{l}}} \over 2} 
 {{1-(-1)^{N_{1}^{l}+n_{l}}} \over 2}.
\end{eqnarray}
This factor takes the values 0 or 1 and its presence, using the fact that in Eq. (\ref{PRODUCT1}) the 
function $\chi( {\cal C}_{1})$ occurs, involves a third characteristic function $\chi({\cal C}_{3})$ 
which accounts for the fact that
if for some $m,\ n_{m}=0$, then there must be one and only one $j$ such that $N_{1}^{mj}=1$. Analogously, if for
some $m,\ \nb_{m}=0$, then there must be one and only one $k$ such that $N_{1}^{km}=1$. 
If now we use also the function $\chi( {\cal C}_{2})$, 
then for any given $\vb{s}$ (\ref{esse}) we find that the set of the 
effective trajectories ${\cal C}$ is given by: 
\begin{eqnarray}
{\cal C} &=& \{
 \mbox{set of trajectories such that:}
 \\ && \nonumber
 \mbox{if $\nb_{i}=1$, any process on the row $i$ of $\vb{\cal N}$ does not jump};
 \\ && \nonumber
 \mbox{if $n_{j}=1$, any process on the column $j$ of $\vb{\cal N}$ does not jump};
 \\ && \nonumber
 \mbox{if $\nb_{i}=0$ and $n_{j}=0$, there is one and only one process 
jumping to 1} 
 \\ && \nonumber
 \mbox{on the $i$th row and on the $j$th column of $\vb{\cal N}$}
\} .
\end{eqnarray}
Clearly if $\bar{k} \neq k$ then ${\cal C}$ reduces to the null set and the integral
 (\ref{NUMERATORTILDE}) gives 0. From now on, we will always suppose 
the $\bar{k}=k$ case. 

From its definition we see that ${\cal C}$ has $(g-k)!$ elements so
that we can represent it in terms of
suitable effective matrices $\vb{\cal N}$ in the following way.
First we must erase all the rows with label $i$ if $\nb_{i}=1$ and the columns
with label $j$ if $n_{j}=1$. Then we must consider all the ways to fill the remaining
$(g-k) \times (g-k)$ sub-matrix with $(g-k)$ times ``1'' in an array  
such that two of them are never in the same row or in the same column,
like for a determinant. 

Let us consider as an example the case $g=5$ and $k=2$ with  
$n_{1}=n_{2}=n_{3}=\nb_{1}=\nb_{2}=\nb_{3}=0$ and $n_{4}=n_{5}=\nb_{4}=\nb_{5}=1$.
In this case ${\cal C}$ has $6$ elements represented by $6$ matrices $\vb{\cal N}$, like e. g. 
(the symbol ``$\times$'' means that the corresponding process never jump)
\begin{eqnarray*}
 \vb{\cal N}=\left(
\begin{array}{ccccc}
 0 & 1 & 0 & \times & \times \\
 1 & 0 & 0 & \times & \times \\
 0 & 0 & 1 & \times & \times \\
 \times & \times & \times & \times & \times \\
 \times & \times & \times & \times & \times \\
\end{array}
\right).
\end{eqnarray*}

Up to now we have not spoken about the event in which no process jumps (with probability $e^{- | \Gamma |}$). It is immediate
to observe that such an event is effective only if $n_{i}=\nb_{i}=1$ for any  $i$ from 1 to $g$ and that the 
integral (\ref{NUMERATORTILDE}) gives simply 1. So, from now on, we will ignore this
trivial case.
\subsection{Phases}

Let us face now the problem of finding a closed expression for the ${\cal S}_{ij}$ 
(see Eq. (\ref{PRODUCT1})):
\begin{eqnarray}
{\cal S}_{ij}=   \exp \biggl[ \int_{[0,1)} i \pi [g+i+j+1+ \sum_{m=i}^{g}(\Nb_{s}^{m}+\nb_{m})+
\sum_{m=1}^{j-1}(N_{s}^{m}+n_{m}) dN_{s}^{ij} \biggr].
\end{eqnarray}
Then we will need to consider the product on the couples $(ij)$ for every $(ij)$ such that $N_{1}^{ij}=1$.
Let us factor the rhs of the above equation in the following form:
\begin{eqnarray}
&&{\cal S}_{ij}=   \exp \biggl[ i \pi [g+i+j+1] N_{1}^{ij} \biggr] \nonumber \\ && \times
\exp \biggl[ \int_{[0,1)} i \pi  [ \sum_{m=1}^{j-1}(N_{s}^{m}+n_{m})+\sum_{m=i}^{g}(\Nb_{s}^{m}+\nb_{m}) ] dN_{s}^{ij} \biggr].
\end{eqnarray}
Let $s^{ij},s^{lm},s^{ml}$ be generic jump times with their corresponding labels, then from (\ref{N&N}) 
we obtain:
\begin{eqnarray}
&&{\cal S}_{ij}=   \exp \biggl[ i \pi [g+i+j+1] N_{1}^{ij} \biggr] \times (-1)^{F_{ij}}
\end{eqnarray}
where
\begin{eqnarray}
&&F_{ij}=[ \sum_{m=i}^{g}\sum_{l=1}^{g}\theta(s^{ij}-s^{ml})N_{1}^{ml}+\sum_{m=1}^{j-1}\sum_{l=1}^{g}\theta(s^{ij}-s^{lm})N_{1}^{lm} ]N_{1}^{ij}+ \nonumber \\ && 
+[\sum_{m=i}^{g} \nb_{m} + \sum_{m=1}^{j-1} n_{m}]N_{1}^{ij},
\end{eqnarray}
which, taking account the definition of ${\cal C}_{1}$, becomes:
\begin{eqnarray}
&&F_{ij}=[ \sum_{m=i+1}^{g}\sum_{l=1}^{g}\theta(s^{ij}-s^{ml})N_{1}^{ml}+\sum_{m=1}^{j-1}\sum_{l=1}^{g}\theta(s^{ij}-s^{lm})N_{1}^{lm} ]N_{1}^{ij}+ \nonumber \\ && 
+[\sum_{m=i+1}^{g} \nb_{m} + \sum_{m=1}^{j-1} n_{m}]N_{1}^{ij}.
\end{eqnarray}
Now in order to calculate the expectation, we must integrate over the jump times and then to sum
over all the trajectories the following product:
\begin{eqnarray}
\label{PRODUCT2}
\prod_{ij}(-A_{ij})\chi({\cal C})\exp \biggl[ i \pi [g+i+j+1] N_{1}^{ij} \biggr] \times (-1)^{F_{ij}}.
\end{eqnarray}
The integration must be done using the probability of the Poisson processes and it is easy to see that
\begin{eqnarray}
 dP=e^{-|\Gamma|}\prod_{ij:N_{1}^{ij}=1}ds^{ij}.
\end{eqnarray}
We should integrate (\ref{PRODUCT2}) with the measure $dP$ and then sum over all the possible events
belonging to the set ${\cal C}$; we should then calculate the following integrals: 
\begin{eqnarray}
\label{INTEGRAFASE}
\int_{0}^{1}e^{-|\Gamma|}\prod_{ij:N_{1}^{ij}=1}(-1)^{H_{ij}}ds^{ij}, 
\end{eqnarray}
where
\begin{eqnarray}
H_{ij}=[ \sum_{m=i+1}^{g}\sum_{l=1}^{g}\theta(s^{ij}-s^{ml})N_{1}^{ml}+\sum_{m=1}^{j-1}\sum_{l=1}^{g}\theta(s^{ij}-s^{lm})N_{1}^{lm} ].
\nonumber
\end{eqnarray}
In appendix D it is proved that 
\begin{eqnarray}
\label{HIJ}
H_{ij}=[ \sum_{m=i+1}^{g}\sum_{l=1}^{g}\theta_{ml}^{ij}N_{1}^{ml}+\sum_{m=1}^{j-1}\sum_{l=1}^{g}\theta_{lm}^{ij}N_{1}^{lm} ],
\end{eqnarray}
where the variables 
with four labels $\theta_{hk}^{ij}$ satisfy the following properties:
\begin{eqnarray}
\label{THETA}
 \theta_{hk}^{ij}+\theta_{ij}^{hk}=1  \nonumber \\
 \theta_{hk}^{ij}+\theta_{hk}^{ij}=0. 
\end{eqnarray}
Now we have the following formula for the integral $I$ (the numerator of (\ref{GAUSS}))
\begin{eqnarray}
 I &=& (-1)^{Q}\Delta_{2g}(-(-1^{\widetilde{s}}))\sum_{N_{1}^{11}=0}^{1}...
\sum_{N_{1}^{gg}=0}^{1} \chi({\cal C}) \prod_{ij:\ N_{1}^{ij}=1}(-A_{ij})
(-1)^{g+1} \nonumber \\
 && \times (-1)^{\sum_{m=i+1}^{g}\nb_{m}+\sum_{m=1}^{j-1}n_{m}}(-1)^{i+j}(-1)^{H_{ij}},
\end{eqnarray}
where $\widetilde{\vb{s}}=(\vb{\nb,n})$ and $(-1)^{Q}$ is the sign necessary to have $I$ from 
$\widetilde{I}$.

Now we will face the problem to obtain a closed expression for each of the
factors appearing in the above equation for $I$.
\subsubsection{Correcting sign $(-1)^{Q}$}
To calculate $(-1)^{Q}$ 
we must consider the exchanges needed to restore the monomial $\xb_{1}^{\nb_{1}}\ldots
 \xb_{g}^{\nb_{g}}\x_{1}^{n_{1}}\ldots \x_{g}^{n_{g}}$  and the ordering of integrations
in the original order. The first produces a sign equals to:
\begin{eqnarray}
 (-1)^{n_{1}(\nb_{1}+\nb_{2}+...+\nb_{g})+n_{2}(\nb_{2}+...+\nb_{g})+...+n_{g}
\nb_{g} },
\nonumber
\end{eqnarray}
the second:
\begin{eqnarray}
 (-1)^{g(g+1)/2}.
\nonumber     
\end{eqnarray}
Therefore
\begin{eqnarray}
\label{SIGNARRAY}
 (-1)^{Q}=(-1)^{n_{1}(\nb_{1}+\nb_{2}+...+\nb_{g})+n_{2}(\nb_{2}+...+\nb_{g})+...+n_{g}
\nb_{g} + g(g+1)/2 }.
\end{eqnarray}
\subsubsection{$\Delta_{2g}(-(-1^{\widetilde{s}}))$}
For $\Delta_{2g}(-(-1^{\widetilde{s}}))$ we have:
\begin{eqnarray} 
 && \Delta_{2g}(-(-1^{\widetilde{s}})) = \prod_{l=1}^{g}\biggl(  {1+(-1)^{\nb_{l}} \over 2}+ {1-(-1)^{\nb_{l}} \over 2}
(-1)^{\nb_{1}}\ldots (-1)^{\nb_{l-1}} \biggr) \nonumber \\
 && \times \prod_{l=1}^{g} \biggl(  {1+(-1)^{n_{l}} \over 2}+ {1-(-1)^{n_{l}} \over 2}
(-1)^{\nb_{1}}\ldots (-1)^{\nb_{g}} (-1)^{n_{1}} \ldots (-1)^{n_{l-1}} \biggr)
\nonumber
\end{eqnarray}
from which one easily has:
\begin{eqnarray}
 && \Delta_{2g}(-(-1^{\widetilde{s}})) = \prod_{l=1}^{g}\exp \biggl( (i \pi) \nb_{l}\bigg[ \nb_{1}+\dots
 +\nb_{l-1}+l-1 \bigg] \biggr) \nonumber \\
 && \times \prod_{l=1}^{g}\exp \biggl( (i \pi) n{l}\bigg[ \nb_{1}+\dots +\nb_{g} +n_{1}+\dots +n_{l-1}+g+l-1 \bigg] \biggr). \nonumber
\end{eqnarray}
So we have:
\begin{eqnarray}
\label{SIGNDELTA}
 \Delta_{2g}(-(-1^{\widetilde{s}})) =(-1)^{R},
\end{eqnarray}
where 
\begin{eqnarray}
 R &=& \nb_{2}(\nb_{1}+1)+\nb_{3}(\nb_{1}+\nb_{2}+2)+\dots +\nb_{g}(\nb_{1}+\dots +\nb_{g-1}+g-1) \nonumber
 \\ && + n_{1}(k+g)+n_{2}(k+g+n_{1}+1)+\dots \nonumber 
 \\ && +n_{g}(k+g+n_{1}+\dots +n_{g-1}+g-1), \nonumber
\nonumber
\end{eqnarray}
i.e.:
\begin{eqnarray}
\label{R}
 R &=& \nb_{2}(\nb_{1}+1)+\nb_{3}(\nb_{1}+\nb_{2}+2)+\dots +\nb_{g}(\nb_{1}+\dots +\nb_{g-1}+g-1) \nonumber
 \\ && + n_{2}(n_{1}+1)+\dots +n_{g}(n_{1}+\dots +n_{g-1}+g-1) \nonumber
 \\ && + (k+g)(n_{1}+\dots +n_{g}).
\nonumber
\end{eqnarray}
\subsubsection{Constant factors}
Due to the structure of the set ${\cal C}$, it is easy to observe that 
the following relations hold:
\begin{eqnarray}
\chi({\cal C}) \prod_{ij:\ N_{1}^{ij}=1} (-1)^{\sum_{m=i+1}^{g}\nb_{m}+\sum_{m=1}^{j-1}n_{m}}=(-1)^{S}, 
\nonumber
\end{eqnarray}
with
\begin{eqnarray} 
\label{S}
S &=& (1-\nb_{1})(\nb_{2}+\dots +\nb_{g})+(1-\nb_{2})(\nb_{3}+\dots +\nb_{g})+ \dots (1-\nb_{g-1})(\nb_{g}) +\nonumber \\
 && (1-n_{2})(n_{1})+(1-n_{3})(n_{1}+n_{2}) \dots (1-n_{g})(n_{1}+ \dots n_{g-1})
\end{eqnarray}
and
\begin{eqnarray}
\chi({\cal C}) \prod_{ij:\ N_{1}^{ij}=1} (-1)^{i+j}=(-1)^{T},
\nonumber
\end{eqnarray}
with
\begin{eqnarray}
\label{T}
T &=& \sum_{i}(1-\nb_{i})i+ \sum_{j}(1-n_{j})j.
\end{eqnarray}
Bringing out of the sum the constant factor $(-1)^{g+1}(-1)$, we arrive at
\begin{eqnarray}
\label{IPARTIAL}
 I &=& (-1)^{Q+R+S+T+g(g-k)}\sum_{N_{1}^{11}=0}^{1}...
\sum_{N_{1}^{gg}=0}^{1} \chi({\cal C}) \prod_{ij:\ N_{1}^{ij}=1}(A_{ij}) (-1)^{H_{ij}}.
\end{eqnarray}
Now we must calculate the parity of the sum $Q+R+S+T+g(g-k)$. We will indicate the parity of an integer $n$ with:
\begin{eqnarray}
{\cal P} [ n ] \equiv \mbox{Parity} [n].
\nonumber
\end{eqnarray}
Using the fact that $\sum_{i}^{g}\nb_{i}=\sum_{i}^{g}n_{i}=k$, it is not difficult to get the following relations:
\begin{eqnarray}
 &&{\cal P} \left[ R+S+T \right] = {\cal P} [ (n_{1}+\nb_{1})+(n_{2}+\nb_{2})3+\dots +(n_{g}+\nb_{g})(2g-1) 
\nonumber \\
 &&+\nb_{1}(\nb_{2}+\dots +\nb_{g})+\nb_{2}(\nb_{1}+\nb_{3}+\dots +\nb_{g})+ \dots \nb_{g}(\nb_{1}+\dots +\nb_{g-1}) 
\nonumber \\
 &&+ (\nb_{2}+\dots +\nb_{g})+(\nb_{3}+\dots +\nb_{g})+\dots +(\nb_{g}) \nonumber \\
 &&+ (n_{1})+ (n_{1}+n_{2})+\dots +(n_{1}+\dots +n_{g-1})+k(g-k) ],
\nonumber
\end{eqnarray}
from which one has:
\begin{eqnarray}
 {\cal P}  \left[ R+S+T \right] &=& {\cal P}  [ (n_{1}+\nb_{1})+(n_{2}+\nb_{2})2+\dots +(n_{g}+\nb_{g})g 
 \nonumber \\ && + k(k-1) + k(g-k) + k(g-1) ],
\nonumber
\end{eqnarray}
so, using Eq. (\ref{SIGNARRAY}) we obtain:
\begin{eqnarray}
\label{GLOBALPHASEPARTIAL}
 && {\cal P}  \left[ Q+R+S+T+g(g-k) \right] = \nonumber
 \\ && {\cal P} [
 n_{1}(\nb_{1}+\nb_{2}+...+\nb_{g})+n_{2}(\nb_{2}+...+\nb_{g})+...+n_{g}\nb_{g} \nonumber \\ 
 && + {g(g+1) \over 2} 
 + (n_{1}+\nb_{1})+(n_{2}+\nb_{2})2+\dots +(n_{g}+\nb_{g})g 
  +g(g-k) ].
\end{eqnarray} 
In the next subsection we will extract another constant factor coming
 from the phases $(-1)^{H_{ij}}$ contained in the product (\ref{IPARTIAL}).
\subsubsection{Permutations - completion of the proof}
Let us now investigate the full meaning of the $H_{ij}$. 
Let be $h=g-k$. 
Let ${\cal I}$ and ${\cal J}$ be the sets (of cardinality $h$) of the labels $i$ and $j$
satisfying $\nb_{i}=0$ and $n_{j}=0$ respectively. We will call such sets ``active''.
First, 
let us suppose that ${\cal I}={\cal J}={\{1,\dots h\}}$.
In terms of the matrix $\vb{A}$, we are looking
at the submatrix of $\vb{A}$ obtained considering the first $h$
rows and the first $h$ columns of $\vb{A}$.

In section 4.2 we have seen that each event of the set ${\cal C}$ is a proper realization of 
the matrix $\vb{\cal N}$. Choosing the rows as reference, we can obtain the events of ${\cal C}$
considering for each of the first $h$ row indices $i$, a permutation $\vb{\pi}=(\pi_{1},\dots ,\pi_{h})$ 
such that $N_{1}^{i\pi_{i}}=1$. In other words, 
for any row $i \in {\cal I}$, $\pi_{i}$ tells us which is the corresponding (unique) active
column. 
With these definitions we can carry out the summation over the events by summing over the
permutations:
\begin{eqnarray}
\sum_{N_{1}^{11},\dots ,N_{1}^{gg}=0,1} \chi({\cal C})f(\cdot)=\sum_{\vb{\pi}\in \vb{\Pi}}f(\cdot),  
\nonumber
\end{eqnarray}
where $\vb{\Pi}$ is the set of permutations of $h$ elements.
Let us define also for each of the first $h$ column indices $j$, the corresponding permutation
$\vb{\omega}=(\omega_{1},\dots ,\omega_{h})$ such that $N_{1}^{\omega_{j}j}=1$.
If $\pi$ and $\omega$ are referred to the same event, it is easy to see that one has
 $\omega_{\pi_{j}}=j$ for any $j\in {\cal J}$.
 
Given an effective trajectory, if $\pi$ is the corresponding permutation, one has
\begin{eqnarray}
\sum_{ij: N_{1}^{ij}=1}H_{ij}=\sum_{i=1}^{h} \big[ \sum_{m=i+1}^{h}\theta_{m\pi_{m}}^{i\pi_{i}} +
\sum_{m=1}^{\pi_{i}-1}\theta_{\omega_{m}m}^{i\pi_{i}} \big].
\nonumber
\end{eqnarray}
Let us now expand the r.h.s. of the above expression in the following way:
\begin{eqnarray}
&& \big[ \theta_{2\pi_{2}}^{1\pi_{1}}+\theta_{3\pi_{3}}^{1\pi_{1}}+\dots +\theta_{h\pi_{h}}^{1\pi_{1}} \big] +
\big[ \theta_{\omega_{1}1}^{1\pi_{1}}+\theta_{\omega_{2}2}^{1\pi_{1}}+\dots +
\theta_{\omega_{(\pi_{1}-1)}(\pi_{1}-1)}^{1\pi_{1}}
\big] \nonumber \\
&& + \big[  \theta_{3\pi_{3}}^{2\pi_{2}}+\dots +\theta_{h\pi_{h}}^{2\pi_{2}} \big] +
\big[ \theta_{\omega_{1}1}^{2\pi_{2}}+\theta_{\omega_{2}2}^{2\pi_{2}}+\dots 
+\theta_{\omega_{(\pi_{2}-2)}(\pi_{2}-2)}^{2\pi_{2}}
\big] \nonumber \\ 
&& +................ \nonumber \\
&& +............. \nonumber \\
&& +.......... \nonumber \\
&& +\big[ \theta_{\omega_{1}1}^{h\pi_{h}}+\theta_{\omega_{2}2}^{h\pi_{h}}+\dots +
\theta_{\omega_{(\pi_{h}-h)}(\pi_{h}-h)}^{h\pi_{h}}
\big]. 
\end{eqnarray}
Let us consider the first row of the above expression. If $\pi_{1} > \pi_{2}$,
then in the first row, besides $\theta_{2\pi_{2}}^{1\pi_{1}}$, $\theta_{\omega_{\pi_{2}}\pi_{2}}^{1\pi_{1}}$ 
occurs too and $\omega_{\pi_{2}}=2$. On the other hand if $\pi_{1} < \pi_{2}$ (the equal sign is not possible since $\vb{\pi}$ 
is a permutation) the former pairing for $\theta_{2\pi_{2}}^{1\pi_{1}}$ is not possible in the same
row, therefore the only possible pairing is with a different row where the upper labels for the $\theta's$ are 
all different from ${1\pi_{1}}$. So in this case the pairing will be with $\theta_{1\pi_{1}}^{2\pi_{2}}$. 
Then, using the relations (\ref{THETA}) one has the following dichotomy: $\pi_{1} > \pi_{2}$ $\rightarrow$ $1$
and $\pi_{1} < \pi_{2}$ $\rightarrow$ $0$. 
Repeating the same argument for any row of the above expression one arrives at:
\begin{eqnarray}
\label{EXCHANGES}
\sum_{ij: N_{1}^{ij}=1}H_{ij} &=& \theta(\pi_{2}-\pi_{1})+\theta(\pi_{3}-\pi_{1})+\dots +\theta(\pi_{h}-\pi_{1})
\nonumber \\ 
&& +\theta(\pi_{3}-\pi_{2})+\dots +\theta(\pi_{h}-\pi_{2}) \dots \nonumber \\
&& +.......... \nonumber \\
&& +....... \nonumber \\
&& +.... \nonumber \\
&& +\theta(\pi_{h}-\pi_{h-1}).
\end{eqnarray}
The r.h.s. of (\ref{EXCHANGES}) is just the number of exchanges needed to bring 
the permutation $\vb{\pi}$ to the inverted fundamental permutation i.e. to the
permutation $(h,h-1, \dots ,1)$. 

If now we want to express the sign of the permutation in more conventional terms, i.e. in
terms of the number of exchanges needed to bring the permutation to the fundamental one
$(1,2, \dots ,h)$, it is enough to consider the relation between such number of exchanges
and the former. If we define $\bar{H}_{ij}$ in the following way:
\begin{eqnarray}
 \sum_{ij: N_{1}^{ij}=1}\bar{H}_{ij} &=& \big[1-\theta(\pi_{2}-\pi_{1})\big]
+\big[1-\theta(\pi_{3}-\pi_{1})\big] +\dots +\big[1-\theta(\pi_{h}-\pi_{1})\big]
\nonumber \\ 
&& +\big[1-\theta(\pi_{3}-\pi_{2})\big] +\dots +\big[1-\theta(\pi_{h}-\pi_{2})\big] \dots \nonumber \\
&& +.......... \nonumber \\
&& +....... \nonumber \\
&& +.... \nonumber \\
&& +\big[1-\theta(\pi_{h}-\pi_{h-1})\big],
\nonumber
\end{eqnarray}
we see that the above r.h.s. now counts the number of exchanges needed to bring the permutation
to the fundamental one; on the other hand we have:
\begin{eqnarray}
\label{PERMUTATION}
\sum_{ij: N_{1}^{ij}=1}H_{ij} = \sum_{ij: N_{1}^{ij}=1}\bar{H}_{ij} - {h(h-1) \over 2}.
\end{eqnarray}
The term ${h(h-1) \over 2}$ is the last of the constant factors just mentioned before. It must be
summed to all the others constant factors, i.e. to the r.h.s. of Eq. (\ref{GLOBALPHASEPARTIAL}). 
So under our hypothesis ${\cal I}={\cal J}={\{1,\dots h\}}$, we have the following final global phase:
\begin{eqnarray}
 && (-1)^{\left[ Q+R+S+T+g(g-k) +{h(h-1) \over 2} \right]} = (-1)^{W_{0}}
\end{eqnarray} 
where $W_{0}$ is given from 
\begin{eqnarray}
 W_{0} &=&  \biggl[
 n_{1}(\nb_{1}+\nb_{2}+...+\nb_{g})+n_{2}(\nb_{2}+...+\nb_{g})+...+n_{g}\nb_{g} \nonumber \\ 
 && + (n_{1}+\nb_{1})+(n_{2}+\nb_{2})2+\dots +(n_{g}+\nb_{g})g 
  +{k(k+1) \over 2} \biggr].
\nonumber
\end{eqnarray} 
On the other hand if ${\cal I}={\cal J}={\{1,\dots h\}}$ it is easy to see that $(-1)^{W_{0}}=1$. \\
Now we can return to the general case in which the active
rows and columns are no more restricted to be the first $h$.
For this aim it is enough to consider the numerator of the (\ref{GAUSS}) and to make a suitable permutation
which brings $d\x_{1}d\xb_{1}...d\x_{g}d\xb_{g}$ in $d\x_{p_{1}}d\xb_{q_{1}}...d\x_{p_{g}}d\xb_{q_{g}}$ with
$n_{p_{1}}=\bar{n}_{q_{1}}=\dots n_{p_{h}}=\bar{n}_{q_{h}}=0$. So we have a again the case ${\cal I}={\cal J}={\{1,\dots h\}}$
up to a sign related to this permutation which is $(-1)^{\sum_{\nu=1}^{h}p_{\nu}+q_{\nu}}
=(-1)^{\sum_{\nu=1}^{k}i_{\nu}+j_{\nu}}=(-1)^{W}$ (see the foonote n.2).
 
Then for the Gaussian, not normalized, Berezin integral $I$ we have found:
\begin{eqnarray}
I=(-1)^{W}\sum_{\vb{\pi} \in \vb{\Pi}}(-1)^{\vb{\pi}}\prod_{i \in {\cal J}} A_{i \pi_{i}},
\nonumber
\end{eqnarray}
where now $\vb{\pi}=(\pi_{1}, \dots ,\pi_{h})$ with $\pi_{i} \in {\cal }J$ for any $i \in {\cal I}$
 and $(-1)^{\vb{\pi}}$ is the
sign of the permutation $\vb{\pi}$, i. e. the sign determined from the parity of the number of exchanges needed
to bring $\vb{\pi}$ to $(j_{1}, \dots ,j_{h})$ with $j_{1}< \dots <j_{h}$. 
So up to the global phase $(-1)^{W}$, $I$ is just the determinant of the $(g-k) \times (g-k)$
submatrix defined in section 2. Then we have found:
\begin{eqnarray}
 I=\det \vb{A_{s^{*}}} (-1)^{W(\vb{s})}  
\nonumber
\end{eqnarray}
\section{Algebraic, geometrical and physical aspects}

Equations (\ref{SOLUZ2}) and (\ref{IDENTITY}) provide 
interesting interpretations and applications in several contests.
\subsection{Algebraic}
If we choose $\vb{s}=(1,1,\dots,1,1)$, Eq. (\ref{IDENTITY}) reduces to the trivial identity
\begin{eqnarray}
\label{TRIVIALIDENTITY}
{\det \vb{C}}={ 1 \over {\det \vb{A}}},
\end{eqnarray}
which tells us, as it is well known, to calculate the determinant of the inverse
of $\vb{A}$, it is not necessary to evaluate the inverse of $\vb{A}$.
Eq. (\ref{IDENTITY}) generalizes this property to any submatrix of $\vb{C}$, $\vb{C_{s}}$, since
to obtain $\det \vb{C_{s}}$ we need just the determinants of $\vb{A}$ and  $\vb{A_{s^{*}}}$, with
$\vb{A_{s^{*}}}$ submatrix of $\vb{A}$. However it is important to note that in general
$\vb{\widetilde{C_{s}}}(-1)^{W}$ is not the inverse of $\vb{A(\widetilde{A_{s^{*}}})^{-1}}$,
where with $\vb{\widetilde{B_{s}}}$ we indicate the $g \times g$ extension of $\vb{B_{s}}$, i.e. 
the matrix representing the operator which is the direct sum 
of the operator in $k$ dimensions, represented by the matrix $\vb{B_{s}}$, plus the identity operator in $g-k$ dimensions.

Actually we can easily prove algebraically Eq. (\ref{IDENTITY}) in cases in which $\vb{A_{s^{*}}}$ is a submatrix of  
$\vb{A}$ with contiguous labels, in the following way.
Let us suppose $\vb{A_{s^{*}}}$ is  the submatrix obtained erasing the last $h$ rows and $h$ columns of $\vb{A}$,
where $h=g-k$. Let us decompose the matrix $\vb{A}$ in four blocks: $\vb{A_{11}}$; $\vb{A_{12}}$; $\vb{A_{21}}$
 and $\vb{A_{22}}$,
with dimensions $k \times k$; $k \times h$; $h \times k$ and $h \times h$ respectively. Let us make the same
for the matrix $\vb{C}$. In particular we have  $\vb{A_{22}}$=$\vb{A_{s^{*}}}$ and $\vb{C_{11}}$=$\vb{C_{s}}$. With
such a decomposition it is easy to multiply any two matrices by considering the blocks like elements of matrices of dimensions $2 \times 2$. Now let us define the matrix 
$\vb{B}$=$\vb{\widetilde{A_{s^{*}}}}+\vb{D}$, where $\vb{D}$ has $\vb{D_{22}}$=$\vb{A_{12}}$ and the others
three blocks zero. Hence one has $\det \vb{B}$=$\det \vb{\widetilde{A_{s^{*}}}} $=$\det \vb{A_{s^{*}}}$. On the other hand 
for $\vb{E}=\vb{CB}$ one has  $\vb{E_{11}}$=$\vb{C_{11}}$; $\vb{E_{12}}$=$\vb{0}$; 
$\vb{E_{21}}$=$\vb{C_{21}}$ and $\vb{E_{22}}$=$\vb{1}$. So, since $\det \vb{E}=\det \vb{C_{s}}$, by using     
\begin{eqnarray}
{\det \vb{A_{s^{*}}} \over {\det \vb{A}}} = \det (\vb{CB}), \nonumber
\end{eqnarray}
we get the identity (\ref{IDENTITY}) (it is simple to see that in this case $W$ is even). 

In the general case
we will have to face the complexity of the indices and the proof of Eq. (\ref{IDENTITY}) can be obtained as a corollary of the Binet - Cauchy's theorem on the compound matrices \cite{AITKEN,GANTMACHER}.

\subsection{Geometrical}
Up to a sign, the determinant of a $g \times g$ matrix represents the volume of a prism in a $g$ dimensional
Euclidean space. The volume is that spanned by $g$ vectors whose $g$ components are the $g$ rows of the matrix.
From Eq. (\ref{IDENTITY}) we have
\begin{eqnarray}
\label{IDENTITY2}
{| \det \vb{A} | \over {| \det \vb{A_{s^{*}}}}|} =\frac{1}{| \det \vb{C_{s}}|}.
\end{eqnarray}
Now, the l.h.s. of this equation is the ratio of two volumes: one is that of a given $g$ dimensional
prism, while the other is that of a $g-k$ dimensional prism obtained by spanning the space using the $g-k$ vectors whose components are given by the $g-k$ rows of $\vb{A_{s^{*}}}$,
i.e. it is a $g-k$ dimensional section of the first prism. 
So this ratio is nothing else than the relative ``height'' with respect to this section (here with ``height'' we mean 
a generalized height as in general it will be itself a lower dimensional volume). 
Hence Eq. (\ref{IDENTITY2}) tell us that $|\det \vb{C_{s}}|$ is the inverse of this height.  
\begin{center}
{
\unitlength=0.500000pt
\begin{picture}(320.00,180.00)(0.00,0.00)
\put(160.00,0.00){\line(2,1){20.00}}
\put(150.00,0.00){\line(2,1){40.00}}
\put(140.00,0.00){\line(2,1){60.00}}
\put(130.00,0.00){\line(2,1){80.00}}
\put(120.00,0.00){\line(2,1){100.00}}
\put(110.00,0.00){\line(2,1){120.00}}
\put(100.00,0.00){\line(2,1){120.00}}
\put(90.00,0.00){\line(2,1){120.00}}
\put(80.00,0.00){\line(2,1){120.00}}
\put(70.00,0.00){\line(2,1){120.00}}
\put(60.00,0.00){\line(2,1){120.00}}
\put(50.00,0.00){\line(2,1){120.00}}
\put(40.00,0.00){\line(2,1){120.00}}
\put(30.00,0.00){\line(2,1){120.00}}
\put(20.00,0.00){\line(2,1){120.00}}
\put(10.00,0.00){\line(2,1){120.00}}
\put(0.00,0.00){\line(2,1){120.00}}
\put(10.00,10.00){\line(2,1){100.00}}
\put(20.00,20.00){\line(2,1){80.00}}
\put(30.00,30.00){\line(2,1){60.00}}
\put(40.00,40.00){\line(2,1){40.00}}
\put(50.00,50.00){\line(2,1){20.00}}
\put(300.00,10.00){\vector(-1,0){100.00}}
\put(320.00,10.00){\makebox(0.00,0.00)[l]{$S=|\det(\vb{A_s}^*)|$}}
\put(320.00,120.00){\makebox(0.00,0.00)[l]{$H=1/|\det(\vb{C_s})|$}}
\put(300.00,60.00){\vector(0,1){120.00}}
\put(300.00,180.00){\vector(0,-1){120.00}}
\put(270.00,180.00){\line(-1,-3){40.00}}
\put(100.00,180.00){\line(-1,-3){40.00}}
\put(100.00,180.00){\line(-1,-1){60.00}}
\put(100.00,180.00){\line(1,0){170.00}}
\put(270.00,180.00){\line(-1,-1){60.00}}
\put(210.00,120.00){\line(-1,-3){40.00}}
\put(40.00,120.00){\line(1,0){170.00}}
\put(0.00,0.00){\line(1,3){40.00}}
\put(60.00,60.00){\line(-1,-1){60.00}}
\put(230.00,60.00){\line(-1,0){170.00}}
\put(170.00,0.00){\line(1,1){60.00}}
\put(0.00,0.00){\line(1,0){170.00}}
\put(160.00,0.00){\line(1,1){60.00}}
\put(150.00,0.00){\line(1,1){60.00}}
\put(140.00,0.00){\line(1,1){60.00}}
\put(130.00,0.00){\line(1,1){60.00}}
\put(120.00,0.00){\line(1,1){60.00}}
\put(110.00,0.00){\line(1,1){60.00}}
\put(100.00,0.00){\line(1,1){60.00}}
\put(90.00,0.00){\line(1,1){60.00}}
\put(80.00,0.00){\line(1,1){60.00}}
\put(70.00,0.00){\line(1,1){60.00}}
\put(60.00,0.00){\line(1,1){60.00}}
\put(50.00,0.00){\line(1,1){60.00}}
\put(40.00,0.00){\line(1,1){60.00}}
\put(30.00,0.00){\line(1,1){60.00}}
\put(20.00,0.00){\line(1,1){60.00}}
\put(10.00,0.00){\line(1,1){60.00}}
\end{picture}}
\end{center}

\subsection{Physical}
\subsubsection{Gaussian averages}
Equations (\ref{SOLUZ1}), (\ref{SOLUZ2}) and (\ref{IDENTITY}) show that the problem of evaluating Gaussian Berezin integrals can be addressed as follows:
\begin{itemize}
\item if $\vb{A}^{-1}$ is known, 
then we will use Eq. (\ref{SOLUZ1}),
\item if $\vb{A}$ is known, 
 then we will use Eq. (\ref{SOLUZ2}),
\item if both $\vb{A}$ and $\vb{A}^{-1}$ are known, we will use Eq. (\ref{SOLUZ1}) if $k \leq g/2$ and Eq.
 (\ref{SOLUZ2}) if $k > g/2$.
\end{itemize}
The last point is obvious.
Since the calculation of a determinant of a $d \times d$ matrix leads to a sum of $d!$ elements with alternate signs, 
the advantage in the choice of one or another representation may be factorial with respect to the dimensions 
as $k$ is enough different from $g/2$.
The others two points, first remarked in \cite{KERLER}, are also obvious. In fact let us suppose we are, for example, in the second case and we are interested
to the not normalized Gaussian integrals (that is, we don't need to evaluate $\det \vb{A}$ for both the two representations).
In order to calculate $\vb{C_{s}}$ from $\vb{A}$ we must calculate $k^{2}$ minors of $\vb{A}$, each of them
in general involving a determinant of a $(g-1) \times (g-1)$ submatrix of $\vb{A}$. Therefore we should to consider
$k^{2}[(g-1)!]$ terms in order to calculate $\det \vb{C_{s}}$, while to calculate $\det 
\vb{A_{s^{*}}}$ we need to consider 
$(g-k)!$ terms. So, in general, for any $k > 1$, knowing $\vb{A}$, the use of 
Eq. (\ref{SOLUZ2}) provides the following gain (calculated as ratio between the number
of terms present in the two representations)
\begin{eqnarray}
\label{GAIN}
\mbox{GAIN}=k^{2}(g-1)(g-2)\dots (g-k+1),
\end{eqnarray}
while for $k=1$ the GAIN is 1.

For normalized Gaussian integrals, actually the gain can be less as the calculation of $\det \vb{A}$
could implicitly provides some of the above $k^2$ minors. However in general the gain will be
of the same order.
\subsubsection{Mapping between two fermionic systems}
Besides the above practical advantage, Eq. (\ref{IDENTITY}) provides an
interesting theoretical insight. 

In terms of Gaussian Berezin averages Eq. (\ref{IDENTITY}) may be read as:
\begin{eqnarray}
\label{IDENTITYAVERAGES0}
<\vb{\psi^{s}}>_{\vb{A}} =
<\vb{\psi^{s^{*}}}>_{\vb{A}^{-1}}\frac{(-1)^{W(\vb{s})}}{\det \vb{A}}.
\end{eqnarray} 
$<\cdot>_{\vb{A}}$ means a Gaussian average being $\vb{A}$ the matrix of the Gaussian action.

Let be given a fermionic system with the following action
\begin{eqnarray}
S(\vb{\x},\vb{\xb})=\frac{1}{2}(\vb{\xb},\vb{A\x})-V(\vb{\x},\vb{\xb}),
\end{eqnarray}
where $V(\vb{\x},\vb{\xb})$ is an arbitrary element of ${\cal G}_{2g}$ which we can expand in monomials as
\begin{eqnarray}
\label{POTENTIAL}
V(\vb{\x},\vb{\xb})=\sum_{\vb{s} \in  ({\mathbb{Z}_{2}}^{*})^{2g}} \vb{\psi^{s}} V_{\vb{s}},
\end{eqnarray}
with $V_{\vb{s}} \in \mathbb{C}$.
The system is then characterized by the following Grassmannian partition function
\begin{eqnarray}
{\cal Z}=\int^{B} \exp [-S(\vb{\x},\vb{\xb})] d\x_{1}d\xb_{1}...d\x_{g}d\xb_{g}. 
\end{eqnarray}

Let $\vb{s^{*}}=(m_{1},\bar{m}_{1};\dots;m_{g},\bar{m}_{g})$. 
Let $\hat{T}$ be a linear operator on ${\cal G}_{2g}$ whose action on monomials $\psi^{\vb{s^{*}}}$ is defined by
\begin{equation}
\hat{T}\psi^{\vb{s^{*}}}=\psi^{\vb{s^{*}}}(-1)^{\sum_{i=1}^{g}[(m_{i}+\bar{m}_{i})i+m_{i}\bar{m}_{i}]+g}
\end{equation}

As shown in appendix A, by using Eq. (\ref{IDENTITYAVERAGES0}), it is not difficult to see that for ${\cal Z}$ we have also
the representation
\begin{eqnarray}
\label{Z}
{\cal Z}=(\det\vb{A})\int^{B} \exp [-S^{*}(\vb{\x},\vb{\xb})] d\x_{1}d\xb_{1}...d\x_{g}d\xb_{g},
\end{eqnarray} 
where now the action $S^{*}$ is given by
\begin{eqnarray}
S^{*}(\vb{\x},\vb{\xb})=\frac{1}{2}(\vb{\xb},\vb{A^{-1}\x})-V^{*}(\vb{\x},\vb{\xb}), 
\end{eqnarray}
and the potential $V^{*}$ is related to $V$ according to the following relationship
\begin{eqnarray}
\label{POTENTIAL*}
V^{*}(\vb{\x},\vb{\xb})= \log \bigg[ \hat{T} \exp [
{V}(\frac{-\vb{\partial}}{\vb{\partial \xb}},\frac{\vb{\partial}}{\vb{\partial \x}}) ] \vb{\psi^{e}}
\bigg], \mbox{ where } \vb{e}=(1,1; \ldots ; 1,1).
\end{eqnarray} 

Let $\vb{\eta},\vb{\bar{\eta}}$ be vectors of ${\cal G}_{g}$ (Grassmann sources).
 If we define the generating Grassmann functional integral
\begin{eqnarray}
{\cal Z}(\vb{\bar{\eta}},\vb{\eta})=\int^{B} \exp [ -S(\vb{\x},\vb{\xb})+(\vb{\eta},\vb{\xb})+
 (\vb{\x},\vb{\bar{\eta}}) ] d\x_{1}d\xb_{1}...d\x_{g}d\xb_{g}, 
\end{eqnarray}
we have also 
\footnote{Let us observe that we adopted for Berezin integration the definition (\ref{BEREZIN}),
which is different from $\int^{B}d\xi_{1} d\xi_{2} \dots d\xi_{g}F(\vb{\xi})=f_{1, \dots , k}$ as $F$,
besides $\xi_{1} \xi_{2} \dots \xi_{g}$, depends on others Grassmann variables too. In this case it is important 
to consider the relations (\ref{FORMALRELATIONS}).}
\begin{eqnarray}
\label{Z1}
{\cal Z}(\vb{\bar{\eta}},\vb{\eta})=(\det\vb{A}) \int^{B} \exp [ 
-S^{*}(\vb{\x},\vb{\xb};\vb{\eta},\vb{\bar{\eta}}) ] d\x_{1}d\xb_{1}...d\x_{g}d\xb_{g}, 
\end{eqnarray}
where now $S^{*}$ is defined as before, but $V^{*}$ depends also on $\vb{\eta}$ and $\vb{\bar{\eta}}$ and is given by
\begin{eqnarray}
\label{POTENTIAL**}
V^{*}(\vb{\x},\vb{\xb};\vb{\bar{\eta}},\vb{\eta})= 
\log \bigg[ \hat{T} \exp [
{V}(\frac{-\vb{\partial}}{\vb{\partial \xb}},\frac{\vb{\partial}}{\vb{\partial \x}}) +
(\vb{\eta},\frac{\vb{\partial}}{\vb{\partial \x}})+(\frac{-\vb{\partial}}{\vb{\partial \xb}},\vb{\bar{\eta}})
] \vb{\psi^{e}} \bigg].
\end{eqnarray} 

Hence we now have, for example, that a fermionic system whose action is represented by 
a free part $\vb{A}$ plus a short range interaction, 
as regard the functional integral, can be viewed as a fermionic system with free part $\vb{A}^{-1}$ plus
a ``long range'' interaction. Substantially $V^{*}$ is the complement of $V$ in ${\cal G}_{2g}$.  
However notice that the term ``long range'' here has not exactly the same meaning used in physics.
The above long range character is related to the fact that if $V$ connects few Grassmann
variables, $V^{*}$ connects many.
\section*{Acknowledgments}
I would like to thank G. Jona-Lasinio and C. Presilla
for their encouragement and many useful discussions and K. Scharnhorst for
critical comments on a previus version of this paper and for bringing  
Ref. \cite{KERLER,CHARRET,KLAUS} to my attention. I would to thank also S. Tanase Nicola for communicating to me his derivation of eq. (\ref{SOLUZ2}) and for interesting comments.
\section*{APPENDIX A}
Let be
\begin{eqnarray}
S(\vb{\x},\vb{\xb})=\frac{1}{2}(\vb{\xb},\vb{A\x})-V(\vb{\x},\vb{\xb}), \nonumber
\end{eqnarray}
a generic action.
Since $(\vb{\xb},\vb{A\x})$ is an even element of $~{\cal G}_{2g}$, we have
$~\exp [-S(\vb{\x},\vb{\xb})]=\exp[-\frac{1}{2}(\vb{\xb},\vb{A\x})]\exp[V(\vb{\x},\vb{\xb})]$ and we can expand the latter
factor in Taylor series with respect to $V(\vb{\x},\vb{\xb})$.
Remembering that $[V(\vb{\x},\vb{\xb})]^{N}$ is finite sum of monomials, we see that 
${\cal Z}$ can be written 
as a finite sum of Gaussian Berezin averages each one multiplied by a suitable coefficient
given by the coefficients of the potential (\ref{POTENTIAL}) and by $\det \vb{A}$ (the Gaussian
normalization). 

Explicitly Eq. (\ref{IDENTITYAVERAGES0}) reads as
\begin{eqnarray}
\label{IDENTITYAVERAGES}
<\x_{1}^{n_{1}}\xb_{1}^{\nb_{1}}\ldots \x_{g}^{n_{g}}\xb_{g}^{\nb_{g}}>_{\vb{A}} =
<\x_{1}^{1-\nb_{1}}\xb_{1}^{1-n_{1}}\ldots \x_{g}^{1-\nb_{g}}\xb_{g}^{1-n_{g}}>_{\vb{A}^{-1}}\frac{(-1)^{W(\vb{s})}}{\det \vb{A}}. \nonumber
\end{eqnarray} 
The r.h.s. of these relations tell us that we can calculate each Gaussian averages considering $\vb{A^{-1}}$ instead 
of $\vb{A}$ as matrix associated to the Gaussian action, provide we use $\vb{s^{*}}$ instead
of $\vb{s}$. $\vb{\psi}^{s^{*}}=\x_{1}^{1-\nb_{1}}\xb_{1}^{1-n_{1}}\ldots \x_{g}^{1-\nb_{g}}\xb_{g}^{1-n_{g}}$ can be obtained
from $\vb{\psi}^{s}=\x_{1}^{n_{1}}\xb_{1}^{\nb_{1}}\ldots \x_{g}^{n_{g}}\xb_{g}^{\nb_{g}}$ by multiple Grassmann derivation applied
to the highest grade monomial $\vb{\psi}^{e}= \x_{1}\xb_{1}...\x_{g}\xb_{g}$:
\begin{eqnarray}
\vb{\psi}^{s^{*}}=\biggl(\frac{\partial}{\partial \x_{1}}\biggr)^{\nb_{1}}\biggl(\frac{-\partial}{\partial \xb_{1}}\biggr)^{n_{1}}
\dots
\biggl(\frac{\partial}{\partial \x_{g}}\biggr)^{\nb_{g}}\biggl(\frac{-\partial}{\partial \xb_{g}}\biggr)^{n_{g}}\vb{\psi}^{e}  
\mbox{,  or} \nonumber
\end{eqnarray}
\begin{eqnarray}
\vb{\psi}^{s^{*}}=(-1)^{n_{1}\nb_{1}+\dots n_{g}\nb_{g}}
\biggl(\frac{-\partial}{\partial \xb_{1}}\biggr)^{n_{1}}\biggl(\frac{\partial}{\partial \x_{1}}\biggr)^{\nb_{1}}
\dots
\biggl(\frac{-\partial}{\partial \xb_{g}}\biggr)^{n_{g}}\biggl(\frac{\partial}{\partial \x_{g}}\biggr)^{\nb_{g}}\vb{\psi}^{e}
\nonumber .
\end{eqnarray}
As regards $(-1)^{W(\vb{s})}$, it can be expressed in terms of $\vb{s^{*}}=(m_{1},\bar{m}_{1};\dots;m_{g},\bar{m}_{g})$ 
and one finds 
\begin{eqnarray}
(-1)^{W(\vb{s}(\vb{s}^{*}))} =(-1)^{W(\vb{s}^{*}(\vb{s}))}=
(-1)^{(m_{1}+\bar{m}_{1})+2(m_{2}+\bar{m}_{2})+\dots g(m_{g}+\bar{m}_{g})}  
\end{eqnarray} 
Hence, we have
\begin{eqnarray}
<V(\vb{\x},\vb{\xb})>_{\vb{A}}=\sum_{\vb{s} \in  ({\mathbb{Z}_{2}}^{*})^{2g}} <\vb{\psi}^{s}>_{\vb{A}} V_{\vb{s}}=
\sum_{\vb{s} \in  ({\mathbb{Z}_{2}}^{*})^{2g}} <\vb{\psi}^{s^{*}}>_{\vb{A}^{-1}} (-1)^{W(\vb{s}^{*}(\vb{s}))} \frac{V_{\vb{s}}}{\det \vb{A}}
\nonumber
\end{eqnarray}
and by using the definition of $\hat{T}$ it is easy to see that one has
\begin{eqnarray}
\label{<V>}
<V(\vb{\x},\vb{\xb})>_{\vb{A}}=\hat{T} 
<{V}(\frac{-\vb{\partial}}{\vb{\partial \xb}},\frac{\vb{\partial}}{\vb{\partial \x}})\vb{\psi}^{e}>_{\vb{A^{-1}}}
\frac{1}{\det \vb{A}}
\end{eqnarray}
and in general
\begin{eqnarray}
\label{<V^N>}
<V^{N}(\vb{\x},\vb{\xb})>_{\vb{A}}=\hat{T} 
<{V}^{N}(\frac{-\vb{\partial}}{\vb{\partial \xb}},\frac{\vb{\partial}}{\vb{\partial \x}})\vb{\psi}^{e}>_{\vb{A^{-1}}}
\frac{1}{\det \vb{A}}.
\end{eqnarray}
Finally, by using Eq.s (\ref{<V>}, \ref{<V^N>}) and the definition of $V^{*}$ (\ref{POTENTIAL*}) 
we get Eq. (\ref{Z}). By using this result the expression (\ref{Z1}) follows as corollary.  
\section*{APPENDIX B}
Concerning $(-1)^{p_{ij}}$, since it accounts of the exchanges 
needed to bring $\xb_{i}$ in front of $\xb_{i}^{{1 \over 2}(1+\eb_{i})}$ and
$\x_{j}$ in front of $\x_{j}^{{1 \over 2}(1+\e_{j})}$ and considering the structure of $\e_{ij}$, 
we see that the first permutation involves $(g-1)+(g-i)$ exchanges and the second one $g-j$. We have
\begin{eqnarray} 
(-1)^{p_{ij}}=(-1)^{g-i-j-1}. \nonumber
\end{eqnarray}
Thus for any given  $\vb{\e} \in{\Gamma}$
\begin{eqnarray}
s(\vb{\e}_{ij})=\Delta_{2g}(\e_{ij})(-1)^{g-i-j-1}A_{ij}. \nonumber
\end{eqnarray}
Let us now calculate $\Delta_{2g}(\e_{ij})$. Due to the arrangements of the generators 
of the algebra, we see from (\ref{DELTA}) that we can break up its expression in one
part which contains only the barred components and another which contains both:
\begin{eqnarray}
&& \Delta_{2g}(\vb{\e}_{ij})=\prod_{l=1}^{g}\biggl(  {1-\eb_{l} \over 2}+ {1+\eb_{l} \over 2}
\eb_{1}\ldots \eb_{l-1} \biggr)\prod_{l=1}^{g} \biggl(  {1-\e_{l} \over 2}+ {1+\e_{l} \over 2}\eb_{1}\ldots \eb_{g} \e_{1} \ldots \e_{l-1} \biggr).
\nonumber
\end{eqnarray}
For the structure of $\vb{\e}_{ij}$ we obtain
\begin{eqnarray}
&& \Delta_{2g}(\vb{\e}_{ij})=(-1)^{g-i} \times (-1)^{g-1+g-j} =  (-1)^{g-i-j-1} \nonumber
\end{eqnarray}
Therefore we finally have
\begin{eqnarray}
s(\vb{\e}_{ij})=A_{ij}. \nonumber
\end{eqnarray}

\section*{APPENDIX C}
Among all the $2g$ factors which are contained in $C_{2g}$, only the $i-$th and the $(g+j)-$th can be different
from $1$, and one has
\begin{eqnarray}
&& C_{2g}(\vb{\e}_{ij},-(-1)^{\widetilde{\vb{N}_{s}}+\widetilde{\vb{n}}})={1+(-1)^{\Nb_{s}^{i}+\nb_{i}} \over 2}
 {1+(-1)^{N_{s}^{j}+n_{j}} \over 2}\nonumber \\ && \times
(-1)^{\Nb_{s}^{i}+...+\Nb_{s}^{g}+N_{s}^{1}+...+N_{s}^{j-1}+\nb_{i}+...+\nb_{g}+n_{1}+...+n_{j-1}}(-1)^{g+j+i+1},
\nonumber
\end{eqnarray}
i.e.
\begin{eqnarray}
&& C_{2g}(\vb{\e}_{ij},-(-1)^{\widetilde{\vb{N}_{s}}+\widetilde{\vb{n}}})={1+(-1)^{\Nb_{s}^{i}+\nb_{i}} \over 2}
 {1+(-1)^{N_{s}^{j}+n_{j}} \over 2} (-1)^{g+j+i+1} \nonumber \\ && \times
\exp \{i \pi [{\Nb_{s}^{i}+...+\Nb_{s}^{g}+N_{s}^{1}+...+N_{s}^{j-1}+\nb_{i}+...+\nb_{g}+n_{1}+...+n_{j-1}}] \} ,
\nonumber
\end{eqnarray}
where, due the structure of $\vb{\e}_{ij}$ and by using Eqs. (\ref{NROWS}) and ( \ref{NCOLUMNS}), one has
\begin{eqnarray}
 \Nb_{s}^{i}=\sum_{l=1}^{g}N_{s}^{il} \nonumber \\
 N_{s}^{j}=\sum_{l=1}^{g}N_{s}^{lj}. \nonumber
\end{eqnarray}  

\section*{APPENDIX D}
Let us suppose that $N_{1}^{ij}=1$, let us consider in (\ref{INTEGRAFASE}), the factor with
labels $(ij)$ and let us observe the structure of $H_{ij}$. 
Let $N^{hk}_{1}=1$, we must consider four events:
\begin{enumerate}
\item $h<i+1$, $k>j-1$ from which none contribute to $H_{ij}$ from $N^{hk}$
\item $h\geq i+1$, $k>j-1$ from which in $H_{ij}$  appears $\theta(s^{ij}-s^{hk})$
\item $h<i+1$, $k\leq j-1$ from which in  $H_{ij}$ appears $\theta(s^{ij}-s^{hk})$
\item $h\geq i+1$, $k\leq j-1$ from which in $H_{ij}$ appears $\theta(s^{ij}-s^{hk})+\theta(s^{ij}-s^{hk})$.
\end{enumerate}
The case n. 1 gives $(-1)^{0}=1$.
The case n. 2 involves $(-1)^{\theta(s^{ij}-s^{hk})}$; as one considers in (\ref{INTEGRAFASE})
the factor with labels $h,k$ one has 
 $i<h+1$ and $j<k+1$ from which $j\leq k$. But two processes cannot jump in the same row or the same column,
it is effective only $j\leq k-1$ and that means that in $H_{hk}$ one has the analog of case n. 3 where
$i,j \rightarrow h,k$ and $vicecersa$. Then one has $\theta(s^{hk}-s^{ij})$, which together with
 $\theta(s^{ij}-s^{hk})$
coming from the former factor, give us $(-1)^{1}=-1$.
The case n. 3 is exactly analogous to the case n. 2.
The case n. 4 gives $(-1)^{2}=1$.

Therefore the integrals over the jump times are fictitious and it is then convenient to define the variables 
with four labels $\theta_{hk}^{ij}$ satisfying the following properties:
\begin{eqnarray}
 \theta_{hk}^{ij}+\theta_{ij}^{hk}=1  \nonumber \\
 \theta_{hk}^{ij}+\theta_{hk}^{ij}=0. \nonumber
\end{eqnarray}
Hence we will write:
\begin{eqnarray}
H_{ij}=[ \sum_{m=i+1}^{g}\sum_{l=1}^{g}\theta_{ml}^{ij}N_{1}^{ml}+\sum_{m=1}^{j-1}\sum_{l=1}^{g}\theta_{lm}^{ij}N_{1}^{lm} ],
\nonumber
\end{eqnarray}
which does not depend on the jump times.


\begin{thebibliography}{99}

\bibitem{B1} F. A. Berezin, $The\ method\ of\ second\ quantization$ (New York: Academic, 1966).

\bibitem{ZJ} J. Zinn-Justin, $Quantum\ Field\ Theory\ and\ Critical\ Phenomena$ 2nd ed. (Oxford University, 1993).

\bibitem{B2} F. A. Berezin, Russian Math. Surveys {\bf 24}, 1 (1969).

\bibitem{B3} F. A. Berezin, Math. USSR Sbornik {\bf 14}, 47 (1971). 

\bibitem{D} B. De Witt $Supermanifolds$ (Cambridge University Press, 1984).

\bibitem{FS} L. D. Faddeev and A. A. Slavnov $Gauge Fields$ (Reading, MA: Benjamin/Cummings, 1980).

\bibitem{DJS} G. F. De Angelis, G. Jona-Lasinio and V. Sidoravicius,
 J. Phys. A: Math. Gen. {\bf 31}, 289 (1998).

\bibitem{AITKEN} A. C. Aitken $Determinants\ and\ Matrices$, University Mathematical Texts (Oliver and Boyd LTD).

\bibitem{GANTMACHER} F. R. Gantmacher $Theory\ of\ Matrices$ (Moskow Nauka, 1967).

\bibitem{KLAUS} K. Scharnhorst, math-ph/0206006. See in particular the eq. (36). 

\bibitem{KERLER} W. Kerler, Zeitschrift f\"{u}r Physik C - Particles and Fields {\bf 22}, 185 - 188 (1984)

\bibitem{CHARRET} I. C. Charret, S. M. de Souza, M. T. Thomaz, Brazilian Journal of Physics {\bf 26}, 720 - 724 (1996)

\bibitem{SORIN} S. Tanase-Nicola recently proved this relation using a Grassmannian transformation which, formally, is a Grassmannian Fourier transformation (private communication).

\bibitem{JP} M. Beccaria, C. Presilla, G. F. De Angelis, and G. Jona Lasinio, Europhys. Lett. {\bf 48}, 243 (1999).

\end{thebibliography}
\end{document}